**Two-step transient liquid phase bonding of NiTi to Ti-6Al-4V through a NbZrW barrier**

Zhaoxi Cao[1], Samuel Price[1], John P. Reidy[1], Ian McCue[1*]

[1]*Department of Materials Science and Engineering, Northwestern University, Evanston, IL 60208, USA*

**Corresponding author:* ian.mccue@northwestern.edu

## Abstract

Dissimilar joining of NiTi to Ti-6Al-4V (Ti-64) is limited by brittle $Ti_2Ni$ formation and degradation of NiTi functionality. A two-step transient liquid phase (TLP) route was developed in which a refractory NbZrW diffusion barrier converts the incompatible couple into two independently bondable interfaces. The barrier plays a different role for each alloy: a reactive substrate for NiTi, dissolving to form a Ti-rich Ni-Ti-Nb liquid that infiltrates its own grain boundaries and is terminated by selective Ti absorption into the barrier; and an inert substrate for Ti-64, joined through contact melting of a sacrificial Cu foil. Both interfaces are fully dense and free of continuous intermetallic layers. In tension, joints spanning both interfaces began transforming at ~350 MPa, exhibited a stress plateau to 2.5% global strain (consistent with stress-induced transformation of the NiTi half of the gauge) and failed beyond the plateau at 380-410 MPa. Five load-unload cycles to 460 MPa showed stable superelastic loops with minor ratcheting. These results demonstrate that a diffusion barrier enables dissimilar TLP joining of otherwise incompatible alloys.

## 1. Introduction

Joining NiTi and Ti-6Al-4V (Ti-64) is important for aerospace and biomedical applications, which often require adaptive structures and components with local flexibility. This pairing combines the recoverable deformation of NiTi (via the superelastic or shape-memory effect) with the specific strength and corrosion resistance of Ti-64. However, neither alloy is well-suited to conventional joining methods. Ti is highly reactive and susceptible to embrittlement when the atmosphere is not tightly controlled. NiTi is similarly sensitive to impurity pickup, but has an added constraint: its transformation temperature, stress, and recoverable strain capability change with composition, precipitation state, and defect density [1]. These issues are magnified during fusion welding – the most widely used joining method – because it generates steep thermal gradients and local chemistry changes in the solidified and heat-affected zones.

Beyond individual alloy constraints, the NiTi/Ti-64 couple is chemically incompatible. Brittle reaction products (e.g., $Ti_2Ni$) form during interdiffusion, and their joints crack under thermal-expansion mismatch or along hard interfacial layers [2]. As a result, direct joining at elevated temperature is generally avoided in favor of interlayers that either block interdiffusion or redirect the local reaction chemistry. Previous reports have used Nb interlayers to reduce direct NiTi/Ti-64 mixing [3], and a Pd interlayer was able to decrease $Ti_2Ni$ formation while allowing the joint to retain useful superelastic behavior under cycling [4]. However, even the best reported joints either rely on fusion processing or retain measurable intermetallic content, and more recent friction-stir and resistance-spot studies report $Ti_2Ni$-associated brittleness and crack susceptibility, with performance well below that of the base alloys [5].

Transient liquid phase (TLP) bonding offers a route to circumvent these limitations. During TLP, a thin interlayer melts and/or reacts with the substrate to form a liquid at the bonding temperature. The liquid wets both surfaces and then solidifies isothermally as the melting-point-depressant solutes diffuse into the adjacent substrates. Melting is confined to the interlayer, so joints form at a modest homologous temperature, with little or no applied pressure and avoiding the steep thermal gradients in fusion joining [6, 7]. TLP is used commercially for the fabrication and repair of Ni-base superalloy turbine components [7], and reactive NiTi-Nb eutectic liquids have produced some of the highest reported joint efficiencies for NiTi, albeit through eutectic solidification on quenching rather than isothermal solidification [8].

Recent work by the authors on similar joining of NiTi demonstrated that CALPHAD-screened Cu-base and Nb-base interlayers form transient liquids which solidify isothermally and yield dense, nearly intermetallic-free joints while preserving the superelastic response [9]. However, this approach cannot be directly extended to TLP of the dissimilar NiTi/Ti-64 couple because any liquid in simultaneous contact with a Ni-rich and a Ti-rich substrate creates brittle intermetallic phases. From a joining temperature perspective, the transient-liquid routes demonstrated for NiTi typically operate at 1120 °C and above, whereas holding Ti-64 above its β-transus temperature (~995 °C) replaces the wrought α+β microstructure with coarse transformed β. As a result, no single bonding temperature is suitable for both substrates.

Here, we address these issues with a two-step TLP route in which a commercial refractory alloy Cb-752 (Nb-5.3W-2.7Zr, at.%), referred to hereafter as NbZrW, is introduced as a barrier between the substrates. The central hypothesis is that the NbZrW layer interrupts direct Ni-rich/Ti-rich interaction, while separate transient-liquid reactions on the NiTi side and Ti side create dense, isothermally solidified interfaces without a continuous brittle layer across the joint. NiTi is first bonded to NbZrW at 1125 °C, where the barrier itself is the reactive component and no interlayer is added, and then the pre-bonded assembly is joined to Ti-64 at 930 °C with a Cu foil that reacts with the substrate to form a Cu-Ti liquid. CALPHAD calculations were coupled with local elemental analysis to determine the equilibrium phase and aid in mapping transport during the joining process. Tensile, pull-to-failure testing of dissimilar joints revealed stress plateaus at ~350 MPa and 2.5% global strain. Cyclic loading of the joint to 460 MPa showed stable superelastic loops with minor ratcheting (<0.3% residual strain accumulated after 5 cycles).

# 2. Methods

## 2.1. Specimen Acquisition and Preparation

TLP bonded joints were prepared using NiTi and Ti-64 rods (25 mm in diameter and 38 mm in length), and a NbZrW alloy (10 wt.% W, 2.5 wt.% Zr, Nb balance) interlayer disc (25 mm in diameter and 3.5 mm in length), with bond surfaces mechanically ground (SiC papers to 1200 grit) to remove surface oxides and reduce surface roughness. Nb was chosen for its chemical compatibility with both NiTi and Ti-64, and the Cb-752 composition was selected over pure Nb and C103 (Nb-10Hf-1Ti, wt.%) for its higher strength.Ti-64 was purchased from McMaster Carr, superelastic NiTi alloy rod feedstock from Kellogg Research Lab, and NbZrW from Elemental Metals.

Figure 1 illustrates the multi-TLP setup. For the NiTi/NbZrW bond, no interlayer was used, whereas the Ti-64/NbZrW bond used a single 10.0 µm-thick Cu foil purchased from Thermo Fisher Scientific. The TLP assemblies were placed in a RD Webb vacuum furnace and evacuated ($10^{-6}$ torr) prior to heat treatment. The NiTi/NbZrW bond was heat-treated at 1125 °C for 2 h and the Ti-64/NbZrW bond was heat-treated at 930 °C for 2 h; for both chemistries, there was no applied external load, the heating rate was 10 ºC/min and the cooling rate was 5 ºC/min until 600 ºC followed by furnace cooling to room temperature.

## 2.2. Microstructural Characterization and Mechanical Testing

To examine the microstructures after heat treatment, cross-sections were mechanically ground and polished down to 50 nm colloidal silica mixed with 5 vol.% $H_2O_2$. Scanning electron microscopy (SEM, FEI Quanta 650 SEM) was performed on the joint interfaces. Energy dispersive spectroscopy (EDS) was used to analyze chemical compositions, rounded to the nearest integer, and electron backscattered diffraction (EBSD) was used to characterize the crystallographic structures and orientations. Large area microstructure maps were constructed via the combined backscatter electron (CBSE) imaging method reported in [10]. Three backscatter electron images of a given region (collected at 100× and stage tilts of 0°, +2°, and -2°) are overlaid as the red, green, and blue channels to form a composite color image of the microstructure.

Tensile specimens were prepared via wire EDM as a scaled-down ASTM E8/E8M, with a gauge length of 23 mm and the other relevant dimensions scaled to keep the same geometric ratios [11]. Tensile tests of the joints, and as-received NiTi and Ti-64, were performed using an MTS Criterion load frame at room temperature with a crosshead speed of 0.006 mm/s, corresponding to a nominal strain rate of order $10^{-4}$ $sec^{-1}$. For cyclic tensile testing, specimens were loaded from 25 MPa to 460 MPa for five cycles before pulling to failure. In-plane strain mapping was performed using digital image correlation (DIC), and a random speckle pattern was applied to the sample surface using an airbrush. Images were acquired at an acquisition rate of 2 Hz using a 4.2-megapixel camera with a 1x lens. Strains were calculated using VIC-2D. Virtual extensometers were used to extract the nominal strain across the gauge length, and averaged strains in NiTi substrate, joint and Ti-64 substrate regions were extracted to quantify the individual material behavior. The joint region was defined as a 3.8mm length measured from the NbZrW/Ti-64 interface towards the NiTi, and the substrate regions extend from either end of the joint region to the ends of the gauge.

Nanoindentation was conducted with a KLA Instruments iMicro nanoindentation tester equipped with a Berkovich tip, using a 30× 5 grid with 50 µm spacing and a 75 mN indentation force. Hardness and modulus values were calculated using the Oliver-Pharr method and were averaged along the width direction and plotted as a function of distance along the length of the dog bone.

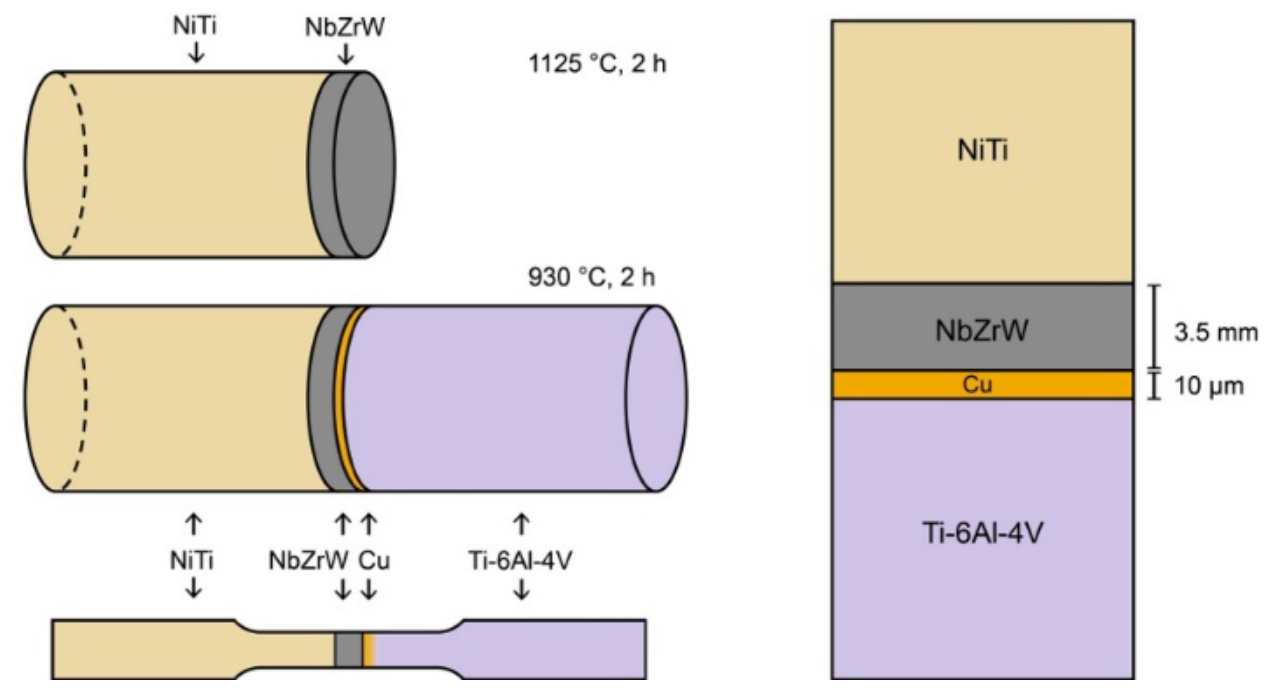

*Figure 1. Schematic of two-step TLP bonding between NiTi-NbZrW and NbZrW-Ti-6Al-4V*

## 2.3. CALPHAD Analysis of Joint Reaction Pathway

The CALPHAD calculations used in this framework were performed by Thermo-Calc software version 2023b using the TC-Python module and TCHEA5 database [12, 13]. Pseudobinary results were obtained by running single equilibrium calculations on a grid of temperature (0-1600°C in increments of 25°C) and composition (100 equally spaced linearly combinations between the two terminal compositions).

## 3. Results and Discussion

### 3.1. Microstructure and Evolution of the NiTi/NbZrW Joint Interface

For the NiTi interface, NbZrW acts as a reactive joining component. Contact with NiTi at 1125 °C generates a transient liquid that penetrates the barrier along its own grain boundaries and solidifies isothermally into a dense joint. The interface after bonding for 2 h is summarized in Figure 2(a-b). The joint is fully dense with no pores or cracks present along the bond line, Figure 2(c-d).

The dominant feature in the BSE images is a second phase, within 250 µm of the interface, decorating the NbZrW grain boundaries. The enclosed NbZrW grains are rounded rather than faceted and the second phase pools at triple junctions, which are characteristic morphologies of boundaries wetted by a liquid rather than solid-state boundary precipitation. Five regions, numbered in Figure 2(b), can be resolved across the joint: (I) the NiTi substrate; (II) a ~150 µm columnar band of NiTi separated from the substrate; (III) a bamboo-grained B2 band at the original NiTi/NbZrW interface; (IV) the grain-boundary penetration zone; and (V) the equiaxed NbZrW matrix.

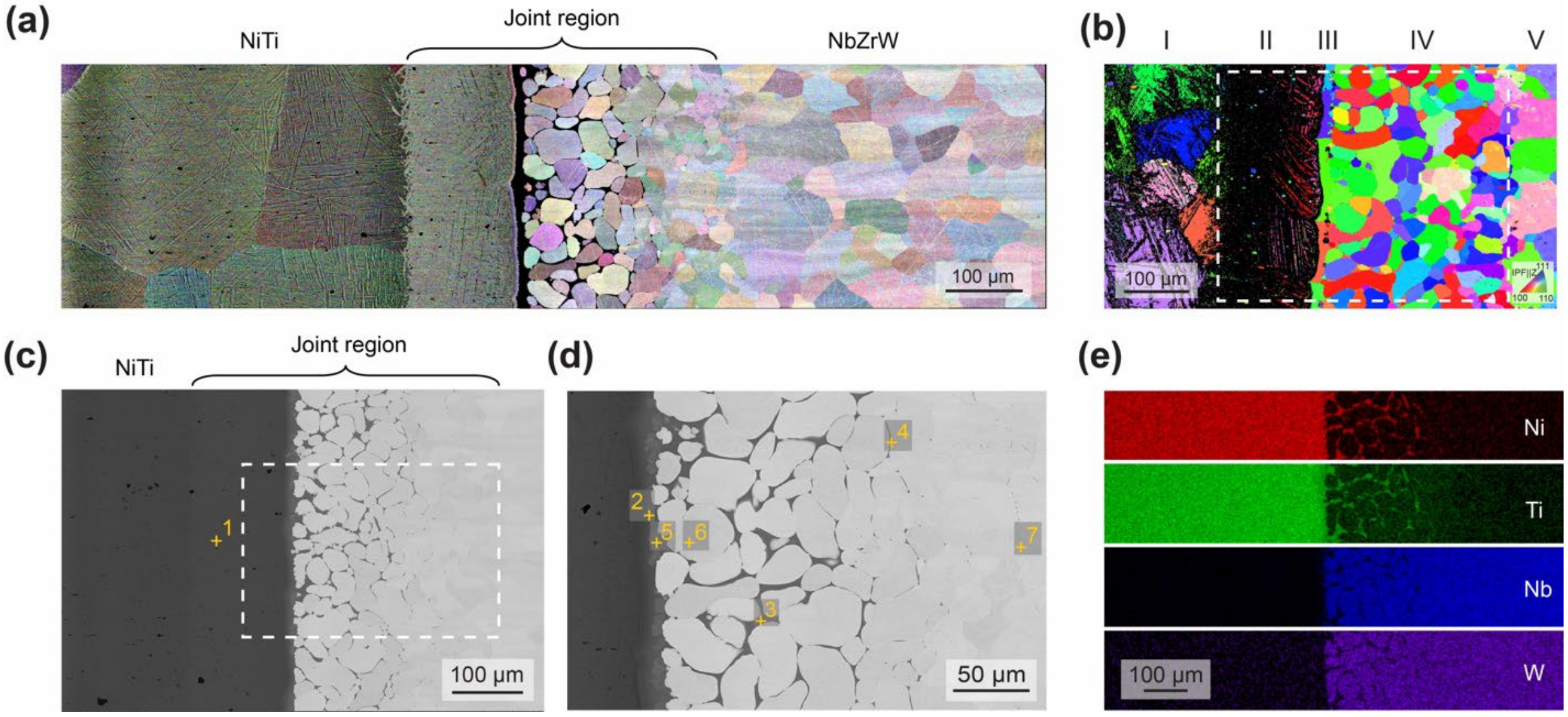

*Figure 2. NiTi/NbZrW interface after bonding at 1125 °C for 2 h. (a) Backscattered-electron orientation-contrast image spanning the NiTi substrate, the joint region, and the NbZrW. Color reflects diode-channel contrast and is not an indexed crystallographic orientation; the NiTi and NbZrW sides were acquired at different detector settings, and the region saturated in the NiTi-side acquisition has been set to black in post-processing, corresponding to the grain-boundary phase and bamboo band resolved in (b). (b) Inverse pole figure (IPF-Z) map of the same interface, resolving the bamboo-grained band at the joint; the dashed white box corresponds to regions II-IV which primarily make up the joint. Numbered regions: (I) NiTi substrate; (II) columnar band; (III) bamboo-grained band; (IV) grain-boundary penetration zone; and (V) NbZrW matrix. Black indicates unindexed points; indexing rates are low on the NiTi side, where mechanical polishing degrades pattern quality. (c, d) Backscattered-electron micrographs of the joint; the dashed box in (c) marks the region shown in (d). Numbered positions correspond to the EDS point analyses in Table 1. (e) EDS maps (Ni, Ti, Nb, and W) across the joint.*

These five regions result from a complex transport process, which was resolved via EDS mapping and point analysis in Figure 2(c–e) and Table 1. Both Ni and Ti penetrated the NbZrW along its grain boundaries (Table 1, points 3 and 4), and Ti is present in the NbZrW grains, decreasing from ~8 at.% at the geometric surface to ~1 at.% at the far edge of the penetration zone (Table 1, points 6 and 7). Ni, by contrast, is at trace levels within those grains but reaches ~50 at.% in the boundary channels. Ti thus leaves the grain boundary liquid while Ni remains during the isothermal hold. Transport into NiTi is negligible; no Nb, W, or Zr is detected above 1 at.% in the NiTi and the Ni content is uniform at ~51 at.% (Table 1, point 1).

*Table 1. Composition of the features identified at the NiTi/NbZrW interface, from EDS point analysis (at.%). Positions are marked in Figure 2(c-d). n.d. ≡ not detected.*

| Position | Feature | Ni | Ti | Nb | Zr | W |
|---|---|---|---|---|---|---|
| 1 | NiTi substrate and columnar band | 51 | 49 | n.d. | n.d. | n.d. |
| 2 | Bamboo-grain B2 band /grain boundary phase at interface | 51 | 44 | 3 | 1 | 1 |
| 3 | Grain boundary phase, mid-penetration | 51 | 36 | 8 | 5 | n.d. |
| 4 | Grain boundary phase, far from the interface | 47 | 28 | 15 | 10 | n.d. |
| 5 | Retained former-liquid pocket | 35 | 44 | 19 | 2 | n.d. |
| 6 | NbZrW grain interior, first grains behind the interface | 1 | 8 | 84 | n.d. | 7 |
| 7 | NbZrW grain interior, far edge of the penetration zone | 1 | 1 | 91 | n.d. | 8 |

The NiTi-based grain boundary phase, Figure 2(c-e), is a B2 structure enriched in Nb (3–15 at.%) and Zr (1–10 at.%); its Ti content falls and its Zr content rises systematically with distance from the interface (Table 1, points 2–4). The bamboo-grained B2 band, region III of Figure 2(b), has the same composition as the grain boundary phase at the interface (Table 1, point 2), consistent with a single continuous phase differing only in grain morphology. The adjacent columnar band, the unindexed region II of Figure 2(b), is NiTi with no resolvable composition gradient and a composition indistinguishable from the base material (Table 1, point 1). Its boundary with the substrate lies parallel to the joint, consistent with a planar diffusion front, and the contrast is attributed to dilute Nb alloying below the EDS detection limit; the microstructure observed in Figure 2(a), but difficult to resolve in Figure 2(b), is likely diffusion-induced recrystallization.

The results from the pseudobinary equilibrium calculations in Figure 3(a-b) reproduce the observed phases and the direction of the measured composition trends; Along a mixing coordinate of mol % NbZrW between NiTi and NbZrW (corresponding to 0 and 100 mol % NbZrW, respectively), a liquid field is stable at 1125 °C between $x \approx 2\%$ and 42%, reaching a maximum mole fraction of 0.34 at $x \approx 20\%$. There is also predicted to be a Zr-rich liquid from 98%-99% NbZrW, however, because the phase fraction of this liquid is ~1% and no experimental evidence of it was found this liquid will not be discussed further. The calculated liquid in Figure 3(e) is Ti-rich and contains no Zr, with a composition range of $Ti_{42–56}Ni_{36–39}Nb_{0–22}W_{0–3.4}$ (at.%), which agrees with the Ti, Ni, and Nb contents of the former-liquid pocket retained at the interface (Table 1, point 5); the 2 at.% Zr in the pocket is assigned to the polygonal precipitates that formed within it on cooling. CALPHAD reproduces the solubility asymmetry measured across the penetration zone: the NbZrW phase, Figure 3(c), is predicted to hold up to ~11 at.% Ti but <1 at.% Ni, and the measured grains contain at most 8 at.% Ti and 1 at.% Ni (Table 1, point 6). The grains are below their calculated Ti solubility at 2 h, which implies diffusion would continue if the heat treatment was longer.

The measured compositions of the grain-boundary phase follow the calculated trend in Figure 3(d) of falling Ti and rising Zr with distance from the interface. The comparison is semi-quantitative – the measured values are ~4 at.% below for Ti and are ~3 at.% above the calculated curves for Ni – but the trend directions are reproduced. Zr is nearly insoluble in both the liquid and the Nb-rich matrix phase, concentrating instead in the NiTi-type B2, Figure 3(d-e), so the Zr enrichment of the boundary channels in Table 1 agrees with these predictions. However, the Zr in the boundary phase is unlikely to have originated from the liquid, due to its insolubility in the liquid, so it likely partitioned directly into the growing B2 at the dissolving matrix interface. Zr is challenging to detect by EDS due to peak overlap with Nb, so while the n.d. entries at points 6 and 7 (Table 1) are consistent with the predicted depletion of the adjacent grains they do not quantify it.

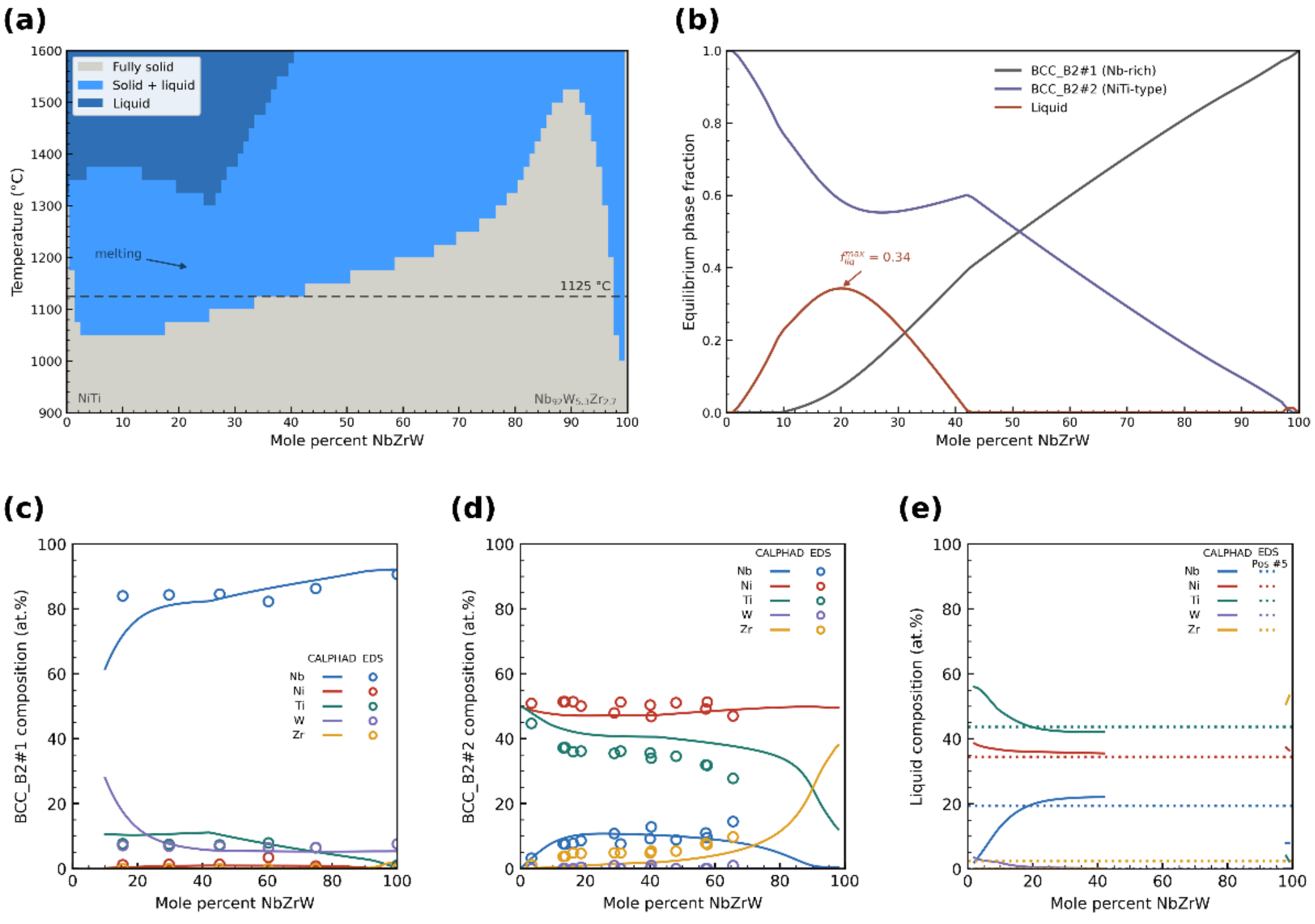

*Figure 3. Equilibrium CALPHAD calculations for the NiTi/NbZrW TLP bonding. (a) Pseudobinary section between NiTi and $Nb_{92}W_{5.3}Zr_{2.7}$, with the 1125 °C bonding isotherm marked showing the regions in which the equilibrium phase are predicted to be solid, liquid, or a combination of both as a function of temperature and a reaction coordinate of mole percent NbZrW. This is used to give insights into the reaction of NiTi and $Nb_{92}W_{5.3}Zr_{2.7}$ during bonding, however, being a pseudobinary diagram the local composition during bonding does not necessarily lie in the plane of this section; panels (b–e) are equilibrium calculations at 1125 °C along the same coordinate and are compared with the measurements semi-quantitatively. (b) Equilibrium phase fractions along the mixing coordinate. Liquid coexists with the NiTi-type B2 (BCC_B2#2) from the onset of melting, reaching a maximum fraction of 0.34 at NbZrW ≈ 20%. (c) Calculated composition of the NbZrW matrix phase (BCC_B2#1) compared with EDS point measurements within the penetration zone. The position of the EDS points along the x-axis is only relative, with the measured distances from the interface being linearly scaled such that furthest measured point from the interface (Table 1, point 7) corresponds to NbZrW=100%. (d) Calculated composition of the grain-boundary phase (BCC_B2#2) compared with EDS point measurements across the penetration zone; the scaling of the x-axis values for the EDS measurements is the same as for (d). (e) Calculated liquid composition compared with measured composition of former-liquid pockets (Table 1, point 5) retained at the interface (dotted lines); since the composition of the former liquid could only be found in isolated pockets the EDS measurement is not given a position on the x-axis and is instead depicted with horizontal lines..*

Combining the experimental and CALPHAD data, bonding is proposed to proceed in three stages. First, solid-state interdiffusion between NiTi and NbZrW raises the local Nb content into the liquid field and a Ti-rich Ni-Ti-Nb liquid forms by contact melting. Melting begins inside a two-phase field, so unmelted B2 coexists with the liquid and melting is partial rather than complete. Region II is the NiTi counterpart of this first stage: interdiffusion continues in the solid state throughout the hold at the Nb-lean edge of the liquid field where the equilibrium liquid fraction is negligible, so the band maintains the substrate composition and is bounded by a planar diffusion front rather than by a solidification front. Second, the liquid wets the NbZrW grain boundaries and is drawn into the alloy along them [14, 15], rounding off the enclosed grains and producing the globular morphology in Figure 2(a,d). Third, the liquid penetration reaches ~250 μm from the interface and solidifies isothermally. The liquid requires a Ti-rich composition (the Ti/Ni ratio is ~1.2 at the maximum liquid fraction), while the NiTi substrate supplies Ti and Ni in near-equal proportion and the NbZrW grains act as a sink for Ti. As a result, the local Ti/Ni ratio falls below that required for the liquid, and the liquid terminates isothermally through selective Ti absorption.

Solidification proceeds by growth of existing B2 and the advancing front remains stable rather than breaking down into cells or dendrites. Each grain of the bamboo-grained B2 band shares the IPF color, and therefore the orientation, of the adjacent grain-boundary B2 channel in region IV, Figure 2(b), so the solidified grains continue the parent structure rather than nucleating. The Nb rejected at the front (the Nb solubility is ~22 at.% in the liquid and ~10 at.% in the B2 solid) is both deposited on the enclosed NbZrW grains, which regrow as the liquid is consumed, and carried in the last liquid, which is retained as the interface pockets (Table 1, point 5, 19 at.% Nb). The bamboo-grained B2 band grows toward the NiTi, and consumption of the last liquid leaves the scalloped boundary between regions III and II in Figure 2(b). Solidification was nearly, but not fully, complete after 2 h. Isolated pockets retained at the interface, Figure S1, contain polygonal precipitates, rather than the single-phase B2 produced by isothermal growth, and the overall composition matches the calculated liquid at the bonding temperature.

This route departs from classical TLP [6, 7] in that NbZrW is the reactive component, and the liquid is generated by partial dissolution of the barrier rather than by melting of a finite interlayer. Owing to these thermodynamic features, there is no finite solute quantity to bound the liquid volume, but the joint extent is instead self-limited by Ti scavenging: the grain-boundary phase at the far edge of the penetration zone holds 28 at.% Ti (Table 1, point 4), well below the ~40 at.% of B2 in equilibrium with the liquid (Figure 3(d)), so the liquid at the far edge had been consumed and the ~250 μm joint region was arrested rather than still advancing at 2 h. The bonding temperature itself sits below the measured NiTi-Nb pseudobinary eutectic of 1151 °C [16], so the binary couple cannot melt at 1125 °C and any liquid requires alloying additions. The role of Zr and W can be estimated by comparison with previous work by the authors on NiTi/Nb bonding under the same conditions [17], where the joint region was confined to 5–10 μm and no grain boundary penetration was observed. The Zr and W additions shift the phase equilibria of the couple such that a liquid is stable at 1125 °C – albeit an intergranular network.

### 3.2. Microstructure and Evolution of the Ti-64/NbZrW Joint Interface

For the Ti-64 interface, NbZrW acts as an inert barrier. The transient liquid forms by contact melting of the Cu foil against the Ti-64, wets the NbZrW face reacting minimally, and solidifies isothermally from the Ti-64 side. This joint, shown in Figure 4 after bonding at 930 °C for 2 h, is fully dense and spans ~165 μm with no pores, cracks, or continuous intermetallic along the bond line. Five regions are resolved in Figure 4(a-b): (I) the coarse-grained NbZrW matrix; (II) a bamboo-grained β band ~40 μm thick along the NbZrW face; (III) an α/β transition band ~50 μm thick, resolved at higher magnification as a nanostructured two-phase mixture (Figure 7) finer than the EBSD step size; (IV) a diffusion-affected band ~75 μm thick of lath α with a Widmanstätten morphology; and (V) the fine equiaxed α+β base microstructure.

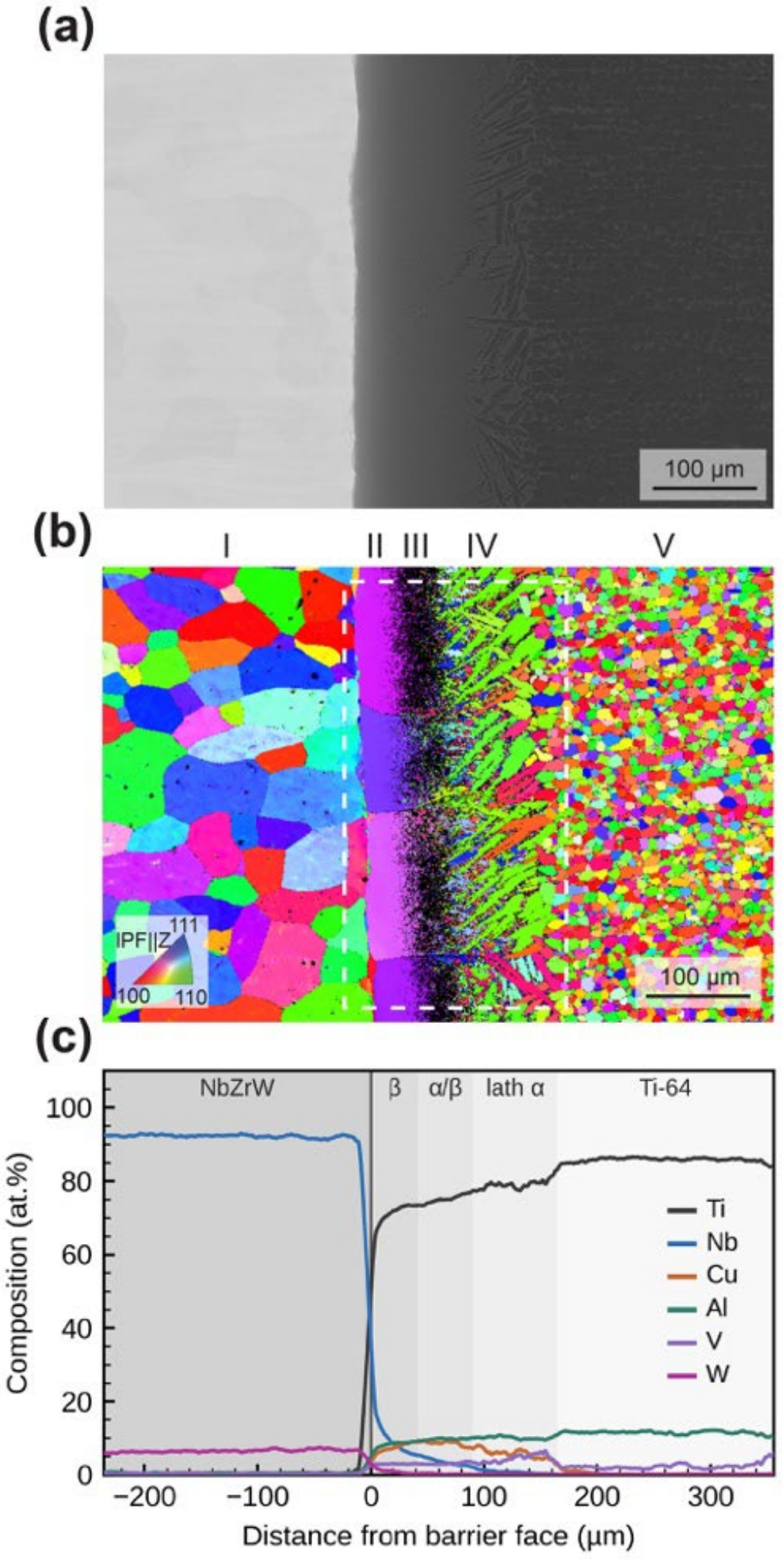

*Figure 4. NbZrW/Ti-64 interface after bonding at 930 °C for 2 h. (a) Backscattered-electron micrographs of the joint. (b) Inverse pole figure (IPF-Z) map of the same interface, resolving the bamboo-grained band at the joint. Five regions are identified and labeled in (c). (c) EDS composition profile across the interface.*

The solute exchange across this interface, using the measured elemental distributions in Figure 4(c), is asymmetric. On the barrier side, Ti uptake is minor, and Cu, Al, and V are effectively zero; Ni is not detected on either side of this interface, justifying the barrier's role to prevent NiTi and Ti-64 from interacting. On the Ti-64 side, Nb steps from its matrix concentration to ~7 at.% at the barrier face and decays to zero over ~100 μm, while Ti steps to ~72 at.% at the barrier face and rises to substrate levels over ~170 μm. Cu is ~5 at.% at the barrier face, peaks at ~8 at.% 50–80 μm from it, and falls to zero at ~165 μm; this 5–8 at.% band marks the distance over which the Cu-containing liquid spread and diffused. Al is diluted compared to its concentration in Ti-64, from 11.3 at.% in the base metal to 9.5 at.% within the band, whereas V is unchanged within measurement uncertainty. W has a short diffusion distance, decaying from 6 to 0 at.% within ~25 μm. Zr was not resolved in the reacted band.

This joint follows the classical TLP bonding of Ti-64 through Cu interlayers [18, 19]. A transient liquid forms by contact melting toward the Cu-rich Ti-Cu eutectic [20] and solidifies isothermally as Cu is diluted into the substrate; 930 °C lies above the lowest liquid-forming eutectic of the Cu-Ti system (~880 °C [20]) and below the β-transus of Ti-64 (~995 °C [21]), so the base α+β microstructure is preserved outside of the diffusion field. The calculated Ti-64/Cu section confirms this behavior: at 930 °C, liquid is stable between mole percent Cu ≈ 27% and 94%, so it exists only where the two have substantially mixed (Figure S3). Although never predicted to be fully liquid at 930 °C, it is nearly so, reaching a phase fraction of 99.6% at ~80% Cu. The barrier, by contrast, has a negligible Cu fraction, reflecting the near immiscibility of Cu and Nb [20], so isothermal solidification is driven entirely from the Ti-64 side, which acts as a sink for Cu. The liquid wets the NbZrW barrier with only limited dissolution: the partial liquid formed on the NbZrW-side of the pseudobinary (Figure S3a) is only in a phase fraction of 3% and the kinetics of dissolution are likely much slower than on the Ti-64 side; the barrier shows no grain-boundary penetration; and the Nb and W carried into the β band correspond to a thin layer of dissolved barrier before isothermal solidification.

Only Ti-64 has appreciable Cu solubility, so the solidification front advances from the Ti-64 side and the last liquid freezes against the NbZrW barrier, whose planar face in Figure 4(b) is the original NbZrW/Ti-64 interface. The reacted band was likely single-phase β during the bonding heat treatment: Cu and Nb are both β stabilizers in Ti-64 [22], and the lath α colonies of region IV, which replace the fine equiaxed α+β of the substrate, indicate that even the outer diffusion field was fully β at 930 °C. The substrate beyond the diffusion field stays α+β because the bonding temperature lies below its β-transus [21]. The phase of each region on cooling is set by its Nb and Cu content and the cooling rate: 5 °C/min is more than two orders of magnitude below the ~20 °C/s at which Ti-64 begins to form martensite [Ahmed and Rack 1998], so β decomposes diffusionally, leaving retained β where Nb and Cu are highest against the barrier, a fine α/β mixture in region III, and Widmanstätten α laths rather than martensite in region IV. The II/III boundary therefore marks where retention ended on cooling, rather than termination of isothermal solidification.

## 3.3. Mechanical Behavior of the NiTi/NbZrW/Ti-64 Joints

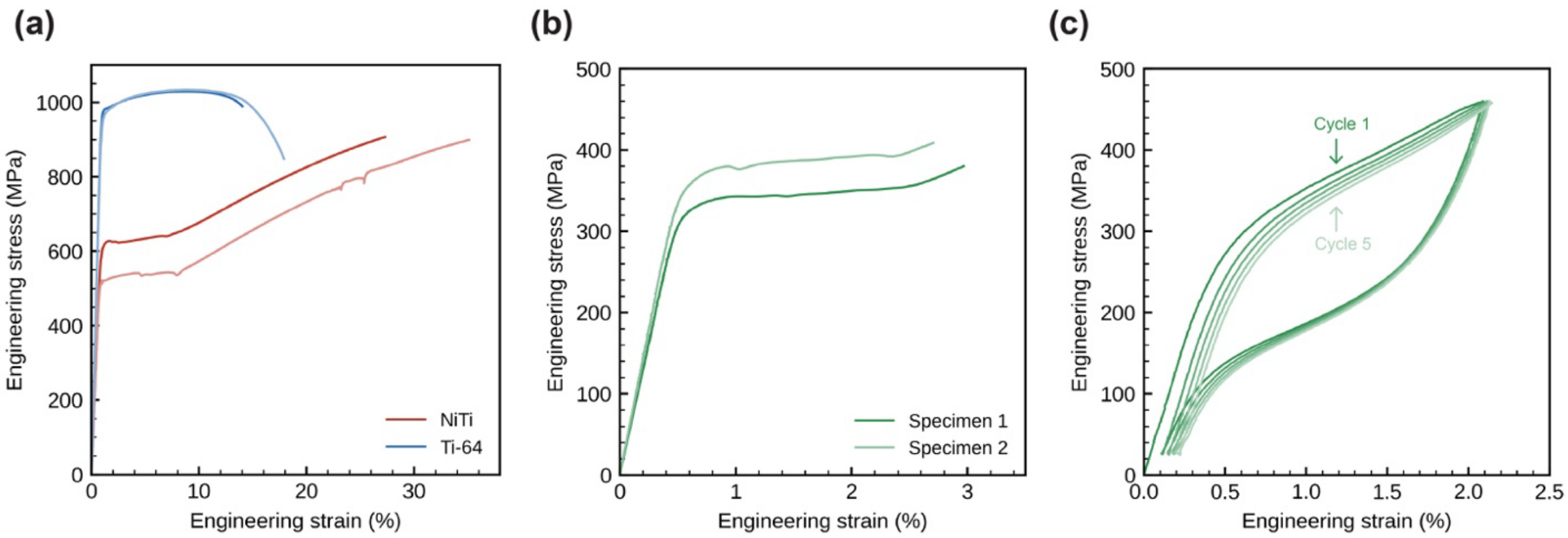

*Figure 5. Room-temperature tensile behavior of the two-step joints and base metals: (a) as-received NiTi and as-received Ti-64; (b) quasistatic response of the joints; (c) cyclic response of the joints (25-460 MPa, five cycles).*

The quasistatic responses of the base metals and joined specimens are compared in Figure 5. The two as-received NiTi specimens transform over stress plateaus at 535 and 625 MPa that extend to 6–10% strain, then harden until failure at ~900 MPa and 27–35% elongation. The two as-received Ti-64 specimens yield at ~960 MPa, reach a UTS of ~1030 MPa, and fail after 14–18% elongation. The two joined specimens deform elastically to ~300 MPa, exhibit transformation plateaus at 350 and 400 MPa beginning at 0.57% and 0.60% strain and then fail at 380 and 410 MPa after ~3.0% elongation. This plateau stress sits ~200 MPa below that of the as-received NiTi, consistent with the removal of cold work and the grain growth produced by the 1125 °C exposure, which lowered the transformation stress of the similar NiTi joints bonded under the same conditions [9]. Only the NiTi half transforms, while the Ti-64 half loads elastically, so the global strain in Figure 5(b) is a diluted measure set by the fraction of the gauge that the NiTi occupies; the plateau ends near 2.5% global strain and the curve hardens to failure at 3% elongation.

The cycled specimen in Figure 5(c) sustained five load-unload cycles to 460 MPa, which is above the two monotonic failure stresses; loading to 460 MPa preceded any cycling, so the survival reflects specimen-to-specimen scatter in joint strength. The specimen showed stable, closed superelastic loops during cyclicing, with a small residual strain accumulating each cycle. The peak strain only grew from 2.10 to 2.14% over the five cycles, while the residual grew from 0.12% after the 1st cycle to 0.23% after the 5th. These strain increments fell from 0.035% to 0.012%, and the DIC analysis below places the residual strain in the NiTi and barrier rather than at either interface.

The monotonic strength of these joints, 380–410 MPa, exceeds that of every interlayered fusion weld reported for this couple except the Pd-interlayered joint. Direct laser welds fail at ~148 MPa and 0.8% elongation [4, 23]; interlayers raise this failure stress to 300 MPa with Cu [24, 25] or Nb [3]; 285 MPa with Co [26]; and 320 MPa at 1.7% strain with Zr [23]. Pd reaches 520 MPa and 5.6% [4], and the joint efficiency here is ~45% of the NiTi ultimate strength and ~40% of that of Ti-64; solid state friction stir by comparison reaches only ~30% of the NiTi UTS [5]. It is difficult to compare cyclic behavior since most of the fusion-based studies only reported monotonic strength [27]; the exception is the Pd-interlayered joint, which exhibited a superelastic plateau at approximately 400 MPa and accumulated <1% permanent strain over 30 load-unload cycles [4]. The TLP route presented here shows comparable cyclic stability over five cycles, with 0.23% residual strain after cycling to 460 MPa, and differs in that the NiTi is never melted, and the interfaces carry no continuous intermetallic layer.

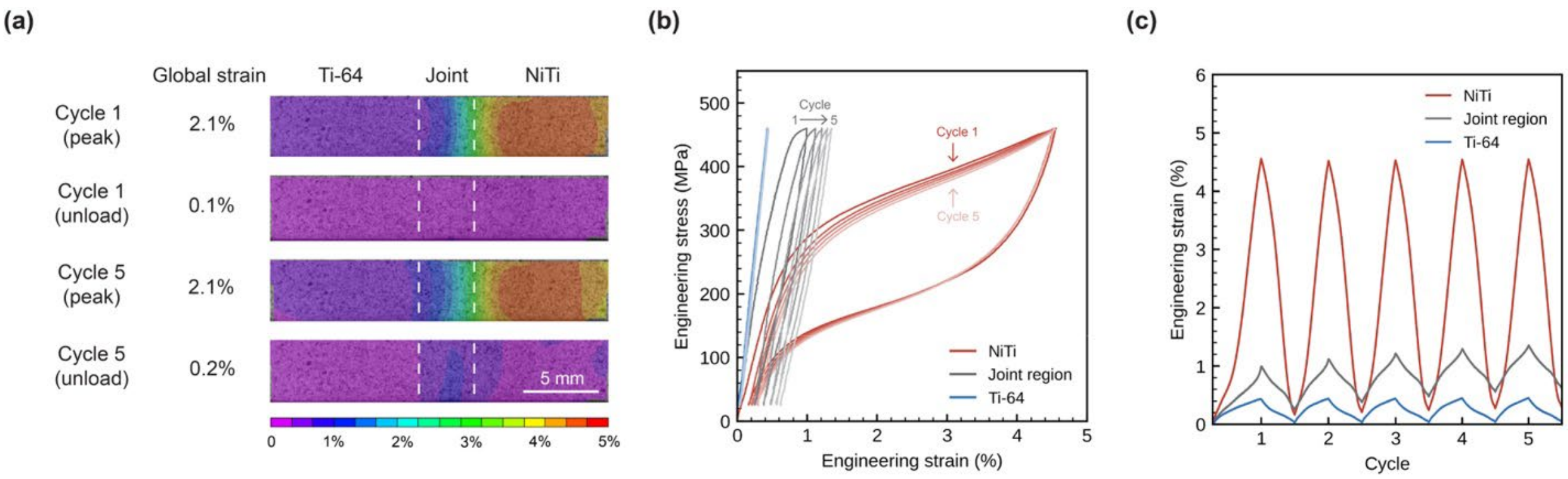

*Figure 6. Digital image correlation analysis of the specimen cycled in Figure 5(c). (a) Axial strain maps at the peak and after unloading of the first and fifth cycles. The dashed lines mark the three virtual extensometer windows used for analysis in (b-c). (b) Local stress-strain response of each region. (c) Strain in each region against cycle number.*

DIC analysis confirms the strain partitioning inferred above. The axial strain maps in Figure 6(a) place the deformation almost entirely in the NiTi half. At the peak of the first cycle, the strain in the NiTi region reaches 4.6% and the joint region reaches 1.0%, while the Ti-64 region is at 0.43% – an order of magnitude below the NiTi. On unloading the strain recovers nearly uniformly to zero, indicating that the NiTi deformation is transformation-dominated rather than plastic. Between cycles 1 and 5, minor strain accumulates within the joint region.

Pairing the applied stress with the strain of each region yields the local stress-strain response of each material in Figure 6(b). The regions are in series, so they carry nominally the same stress and the differences between them are constitutive rather than loading. The NiTi region, loaded above its transformation stress, exhibits the flag-shaped hysteresis loop characteristic of superelasticity; the joint region traces an open but much narrower loop; and the Ti-64 region, far below its yield stress, traces an elastic line. The initial slopes yield 110 GPa for Ti-64, in agreement with literature values (110–114 GPa); an apparent modulus of 45 GPa for NiTi, below the 60–80 GPa of austenitic NiTi, reflecting the early departure from linearity in Figure 6(b); and an apparent modulus of 80 GPa for the joint region, below the ~110 GPa of NbZrW because the joint window spans the 3.5 mm barrier and some of the transforming NiTi, whose compliance dominates.

These differences between regions hold over the five cycles, Figure 6(c). Residual strain accumulates primarily in the NiTi and the joint region, from 0.16% to 0.27% and from 0.25% to 0.55%, respectively. The strain accumulation from each cycle diminishes in these two regions with increasing cycle number, agreeing with the observations from the macroscopic stress-strain curves, while the Ti-64 region stays below 0.07% and returns to its starting length. The global residual of 0.23% is the length-weighted sum of these regional contributions, which is discussed further in the following section.

### 3.4. Role of the NbZrW Barrier in Deformation and Failure

The performance of the three images can be better resolved in Figure 7, which shows the residual strain profile along the gauge after the fifth unload. The joint-region accumulation is twice that of NiTi, with a maximum of 0.58% at the center of the NbZrW region, and the profile falls monotonically to the ~0.27% plateau of the NiTi and to ~0.07% in the Ti-64. NbZrW has no stress-induced phase transformation, so this strain is most likely plastic deformation. The peak cyclic stress is consistent with reported yield strengths for Cb-752 (400-460 MPa [28]), and the joint region's peak strain grew by the same amount as its residual strain, as expected for permanent deformation.

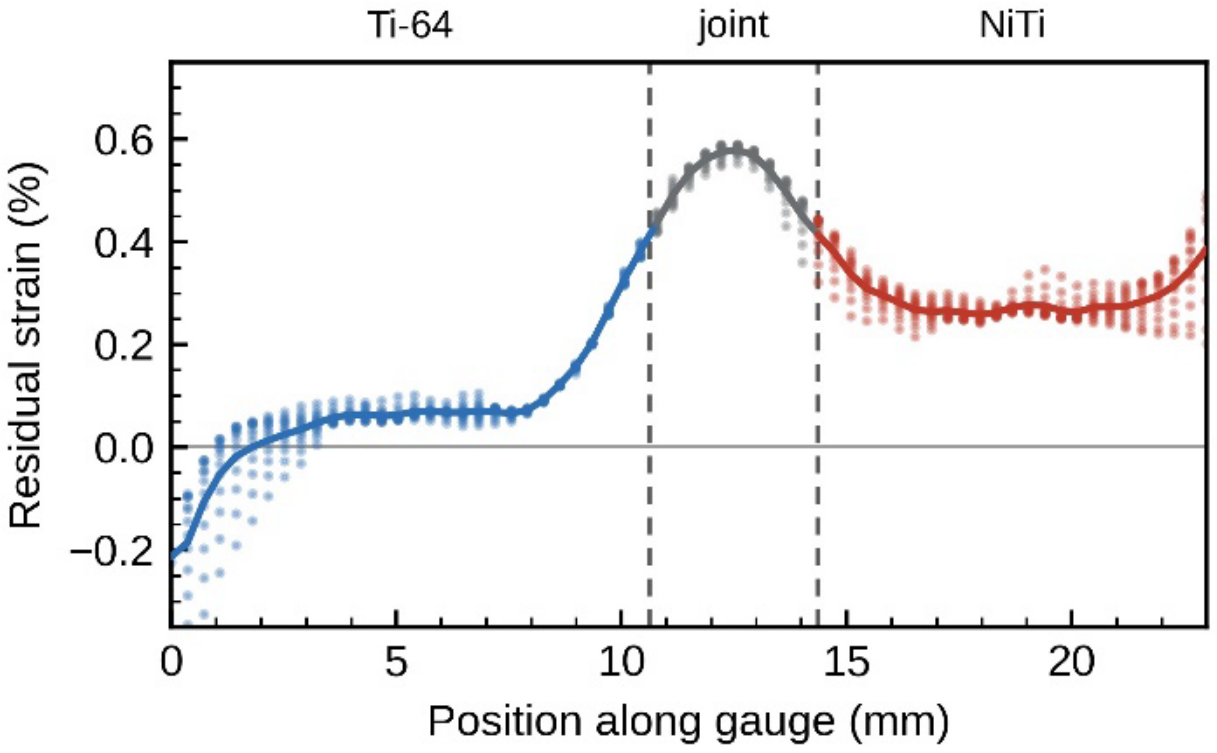

*Figure 7 Residual strain along the gauge length after unloading from the fifth cycle (25–460 MPa), collapsed along its width: (blue) the Ti-64 region, (gray) the joint region, and (red) the NiTi region. Points are individual rows of the DIC strain field and the line is the width-averaged profile. Dashed lines mark the virtual extensometer windows of Figure 6; the negative values within ~1.5 mm of the left end is a grip-end artifact.*

The data in Figure 6 and Figure 7 demonstrate that the functional behavior is set by NiTi, which carries the transformation strain, while Ti-64 remains elastic. No strain concentrations were resolved along the gauge length at the peak cyclic stress, and the residual profile in Figure 7 shows no step at either interface after five cycles. Nevertheless, all three specimens failed at the NbZrW/Ti-64 interface, and the fracture path is resolved by EBSD in Figure 8 for one of the two monotonic specimens. The retained β band remains attached to the Ti-64 half, Figure 8(a), and bare NbZrW grains form the face of the barrier half, Figure 8(b).

The fracture is flat and separates the β band from the barrier, so the crack ran along the planar, geometric interface where the last liquid solidified.

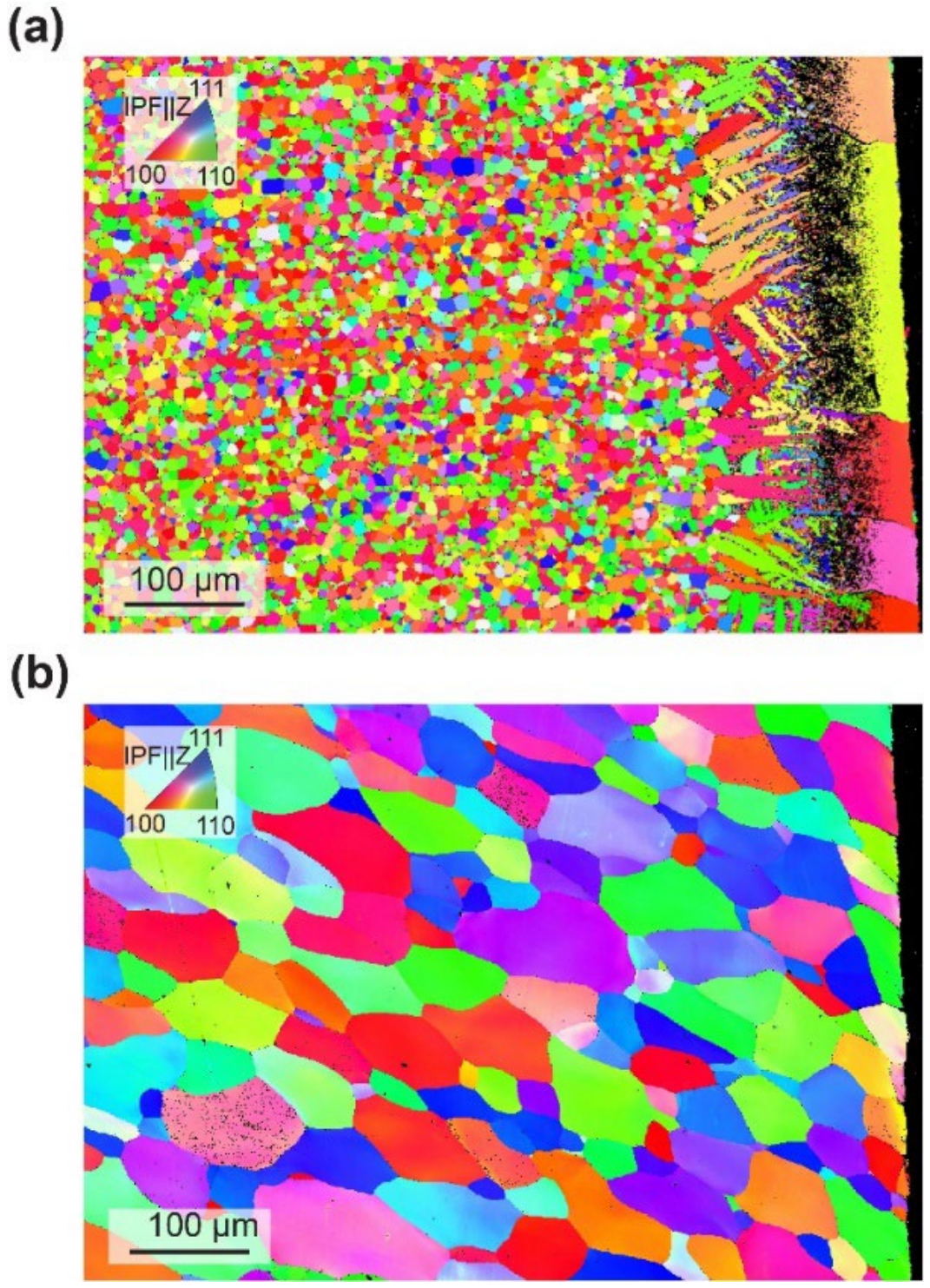

*Figure 8. Inverse pole figure (IPF‖Z) maps of the two halves of the cycled specimen after pull-to-failure: (a) the Ti-64 half and (b) the NbZrW half. The fracture surface is the black region at the right of each panel. The retained β band remains on the Ti-64 half and bare NbZrW grains form the fracture face of the barrier half, so the failure ran along the NbZrW/β-band interface; the unindexed band in (a) is the fine α/β mixture of region III.*

The strength of this interface is set by two key features. Owing to the large mismatch in yield stress, once the NbZrW deforms plastically its lateral contraction exceeds the elastic contraction of the β band and Ti-64, and the resulting shear is carried by the interface; in contrast, the interlocked penetration zone on the NiTi side can distribute the strain over (what is effectively) a composite. Second, the interface is planar, so one failure initiates it can propagate along a straight line of grain boundaries without any impediment. Nanoindentation along the Ti-64/NbZrW interface (Figure S4), with indents every 50 µm, showed no hardness or modulus anomaly within 500 µm, which excludes a hardened or intermetallic layer thicker than the indent spacing; a sub-micrometer segregation or oxide film at the interface could be a possible source of failure but were not observed under scanning electron microscopy.

This characterization reveals several potential strategies to improve the performance of the dissimilar NiTi/Ti-64 joint: (i) a thinner barrier, (ii) a reactive chemistry at the Ti-64 face, and (iii) a stronger barrier. (i) The penetration depth on the NiTi side is ~250 µm and the reacted band on the Ti-64 side is ~165 µm, so a ~1 mm barrier would still confine Ni, and have a smaller contribution (by a factor of 3) to the overall permanent strain in the joint assembly. (ii) Grain-boundary penetration by a liquid is typically avoided as a precursor to liquid-metal embrittlement, but here the penetrating liquid solidified to NiTi-type B2 and left an interlocked zone that survived cycling, whereas the face that was wetted without dissolution failed after modest strain; a Ti-64-specific chemistry that reacts and partially dissolves the barrier surface could produce the same interlocking and potentially impart strength. (iii) Finally, given that the assembly fails due to the substantial mismatch between NbZrW and Ti-64, a barrier with a higher room-temperature yield strength (that survives the 1125 °C exposure) would raise the assembly's failure stress and total elongation, while reducing residual strain and potentially improving cyclic performance.

## 4. Conclusions

In this study, a two-step TLP bonding route was proposed and developed for dissimilar joining of NiTi to Ti-64 through a refractory NbZrW barrier. The NiTi was bonded to NbZrW at 1125 °C for 2 h with no added interlayer, and the assembly was then joined to Ti-64 at 930 °C for 2 h through a 10 µm Cu foil. Both interfaces were fully bonded, and the joints retained a superelastic response under monotonic and cyclic loading. The main conclusions are as follows:

(1) The NbZrW barrier converts the incompatible NiTi/Ti-64 couple into two independently bondable interfaces that are dense with no continuous intermetallic phase.

(2) Each interface forms its transient liquid via contact melting, but by different mechanisms and transport pathways: on the NiTi side, the barrier itself dissolves to form a Ti-rich Ni-Ti-Nb liquid that infiltrates the NbZrW grain boundaries and is terminated by selective Ti absorption into the barrier; on the Ti-64 side, a sacrificial Cu foil forms a Cu-Ti liquid against a barrier that it wets without penetrating and is terminated by Cu dilution into the Ti-64. CALPHAD calculations along the mixing coordinates reproduce the phases and the direction of the composition trends measured at both interfaces.

(3) Both joints solidify isothermally, with the fronts advancing in opposite directions relative to the barrier: away from it on the NiTi side, where a bamboo-grained B2 band grows on the grain-boundary B2 and the last liquid is consumed at the NiTi; and toward it on the Ti-64 side, where the last liquid freezes against the face.

(4) The joints exhibit a transformation plateau beginning near 350 MPa that continue to 2.5% global strain (consistent with stress-induced transformation of the NiTi half of the gauge), fracture at 380–410 MPa, and stable superelastic loops over five load cycles to 460 MPa, with a 0.23% global residual strain shared between transformation training of the NiTi and plastic ratcheting of the barrier.

(5) The strength of the assembly is set by the NbZrW/Ti-64 interface: all three specimens fractured along this interface between 380–410 MPa, without prior strain localization at that interfaceand while the NiTi half underwent substantial cyclic transformation strain.

## 5. Acknowledgements

This work is supported by NASA grant number ECF 80NSSC21K1810. Both S.P and J.P.R received support from the Department of Defense (DoD) through the National Defense Science & Engineering Graduate (NDSEG) Fellowship Program. This material is based upon work supported by the Air Force Office of Scientific Research under award number FA9550-23-F-0014 in the amounts of $139,400 and $143,000. The authors thank Othmane Benafan, Santo Padula, and Travis Turner for the thoughtful discussions and feedback.

This work made use of the EPIC facility of Northwestern University's NUANCE Center, which has received support from the SHyNE Resource (NSF ECCS-2025633), the IIN, and Northwestern's MRSEC program (NSF DMR-2308691); and the MatCI Facility supported by the MRSEC program of the National Science Foundation (DMR-2308691) at the Materials Research Center of Northwestern University.

**Supplemental Information for "Two-step transient liquid phase bonding of NiTi to Ti-6Al-4V through a NbZrW barrier"**

Zhaoxi Cao[1], Samuel Price[1], John P. Reidy[1], Ian McCue[1*]

*[1]Department of Materials Science and Engineering, Northwestern University, Evanston, IL 60208, USA*

**Corresponding author: [ian.mccue@northwestern.edu](mailto:ian.mccue@northwestern.edu)*

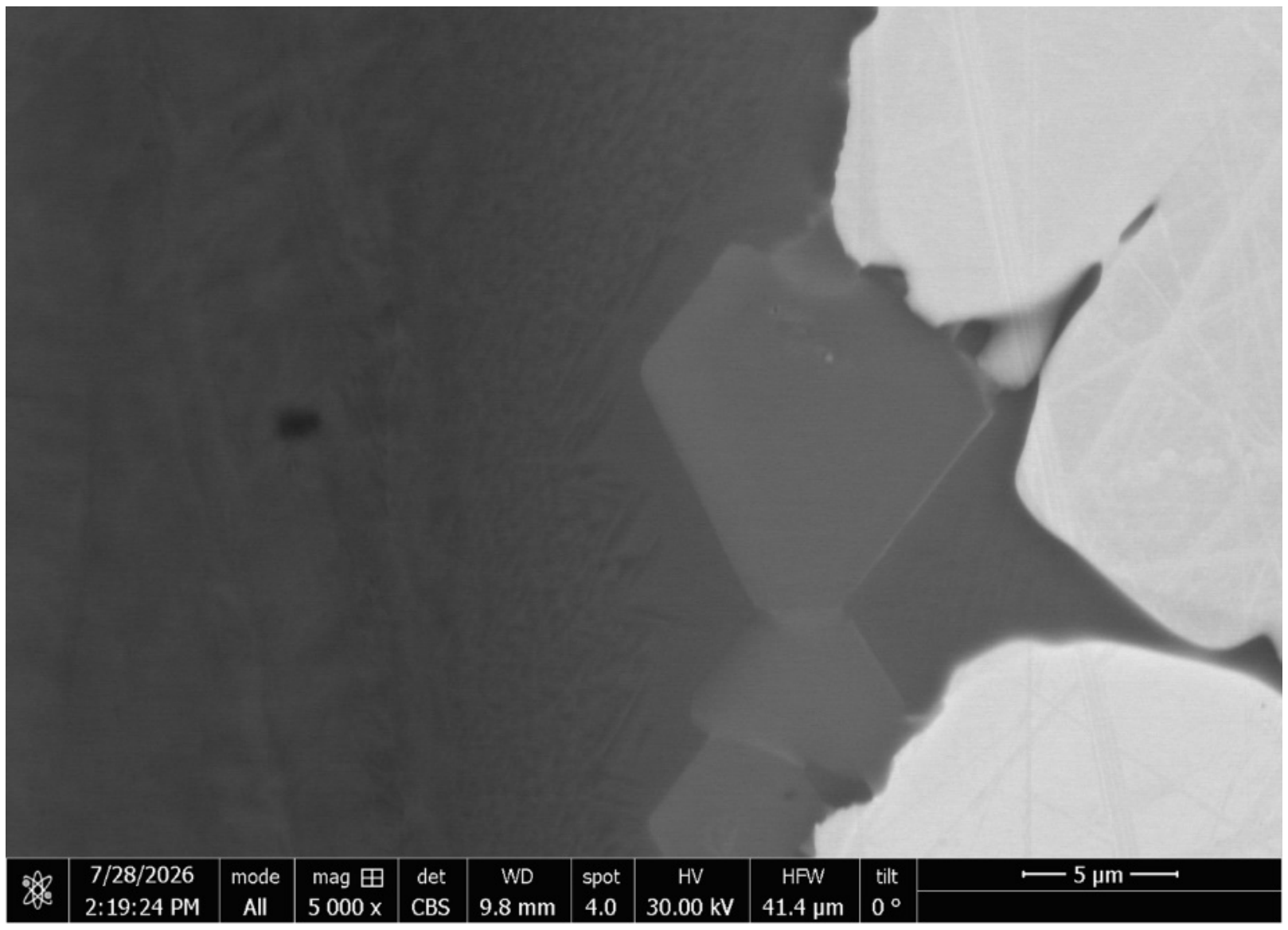

*Figure S1. High-mag BSE image of NiTi/NbZrW interface showing polygonal precipitates, which match the composition of liquid formed during TLP bonding*

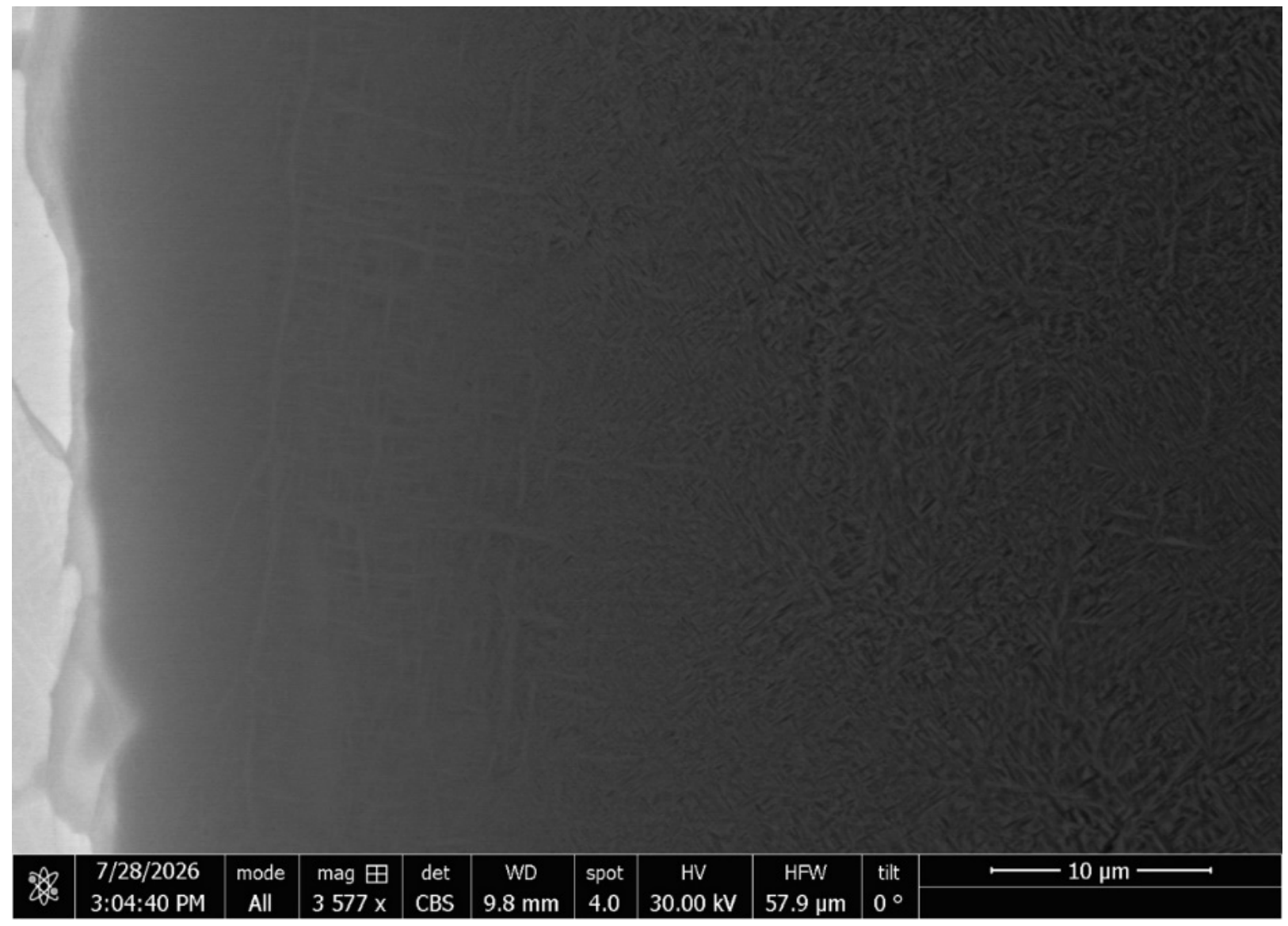

*Figure S2. High-mag BSE image of NbZrW/Ti-64 interface showing nanostructured two-phase microstructure.*

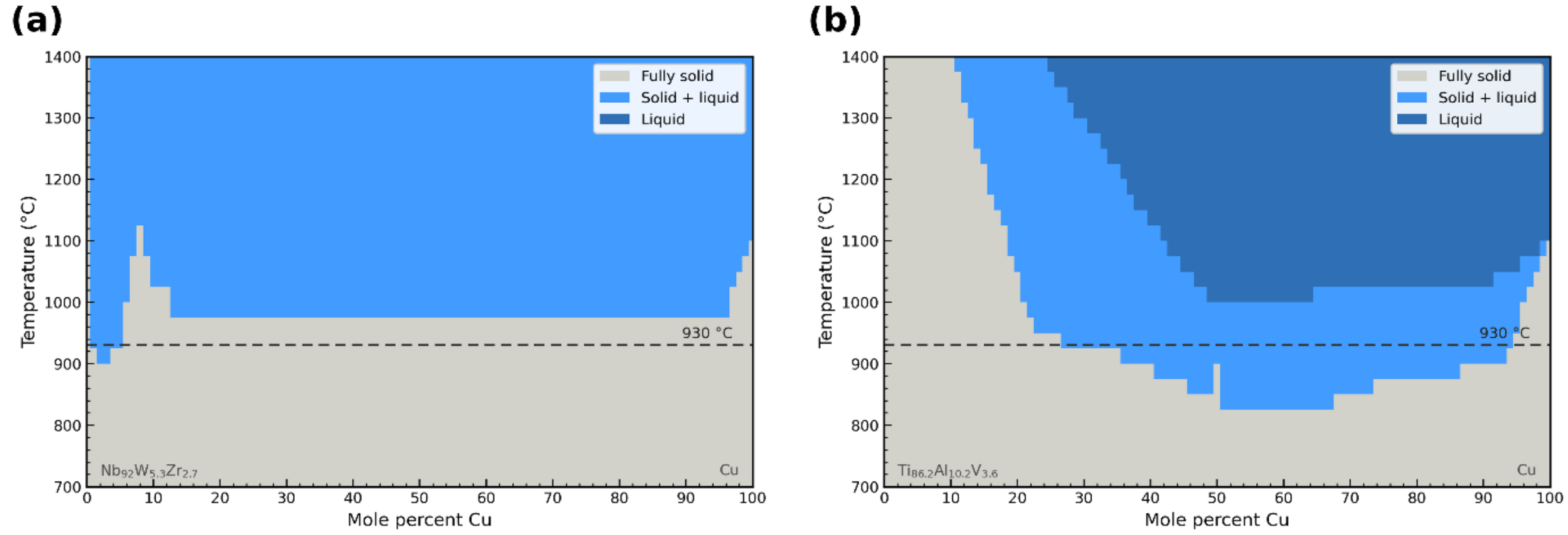

*Figure S3. Pseudobinary section between (a) NbZrW and Cu and (b) Ti-64 and Cu, with the 930 °C bonding isotherm marked. Liquid is stable due to Ti-64/Cu reaction between Cu ≈ 27% and 94%.*

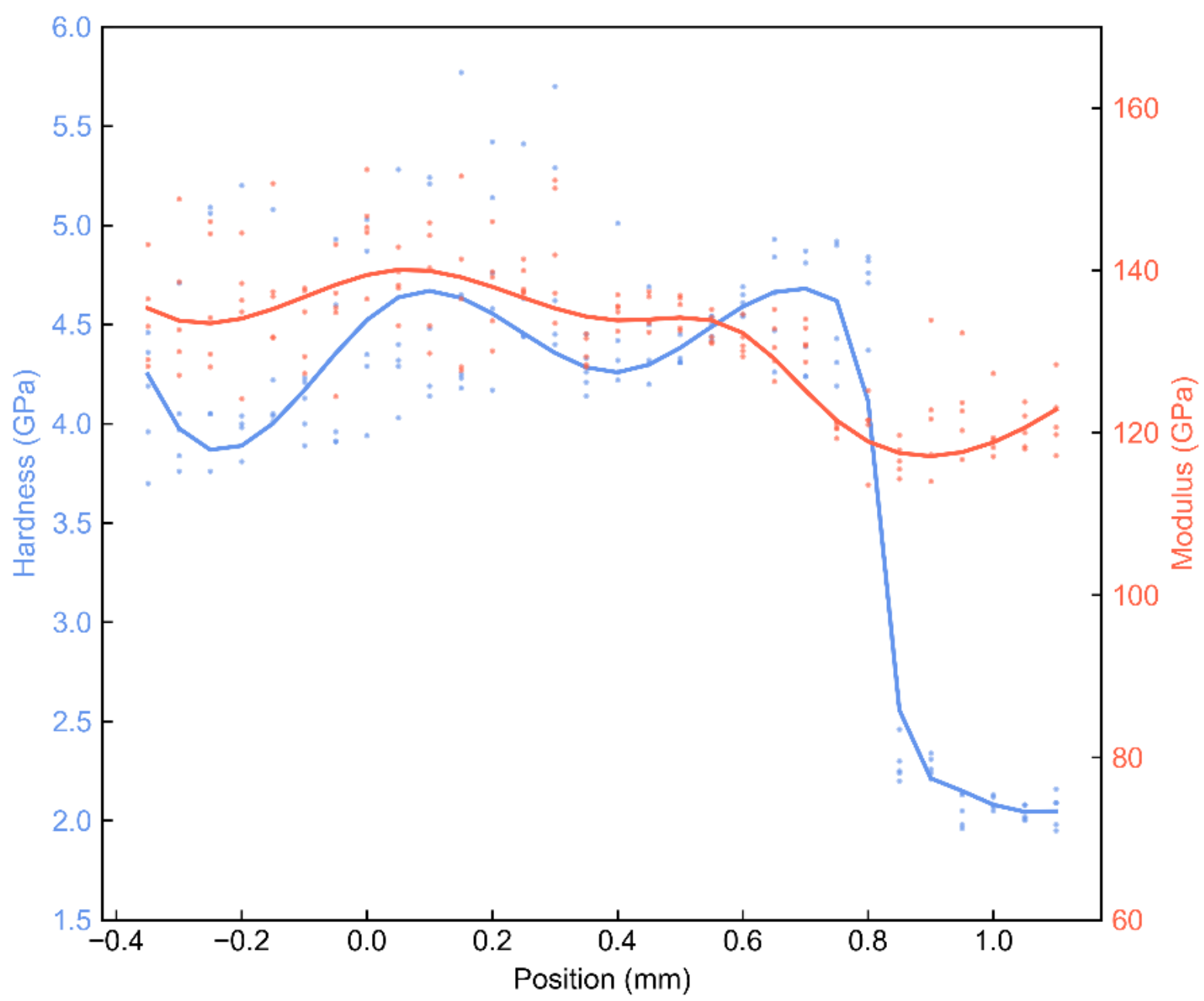

*Figure S4. Nanoindentation mapping of the Ti-64/NbZrW joint. Horizontal lines indicate the modulus and hardness values (averaged over the width direction), and individual data points along the width are plotted as markers.*